\documentclass[twocolumn,showpacs,preprintnumbers,amsmath,amssymb]{revtex4}

\usepackage{graphicx}
\usepackage{dcolumn}
\usepackage{bm}

\begin{document}

\preprint{}

\title{Spectrum of Kinetic Alfv\'en Turbulence}

\author{Stanislav Boldyrev$^1$}
\author{Jean Carlos Perez$^{1,2}$}%
\affiliation{{}$^1$Department of Physics, University of Wisconsin, Madison, WI 53706, USA\\
{}$^2$Department of Physics and Space Science Center, U. New Hampshire, Durham, NH 03824, USA}%
\date{9 February 2012; revised 10 April 2012}

\begin{abstract}
A model for strong kinetic Alfv\'en plasma turbulence at scales smaller than the ion gyroscale is proposed. It is argued that magnetic and density fluctuations are concentrated mostly at two-dimensional structures, which leads to their Fourier energy spectra $E(k_\perp)\propto k_{\perp}^{-8/3}$, where $k_\perp$ is the wave-vector component normal to the strong background magnetic field.  The results are shown to be in good agreement with numerical simulations, and they can explain recent observations of magnetic and density fluctuations in the solar wind at sub-proton scales. 
\end{abstract}

\maketitle

{\em Introduction}---Possibly the simplest description of magnetic plasma turbulence at scales much larger than typical micro-scales (particle gyro-radii, skin depth, etc) is provided by one-fluid  magnetohydrodynamics (MHD), where electrons and ions are assumed to move together as a single fluid \cite[e.g.,][]{biskamp03,kulsrud05}. A characteristic feature of MHD turbulence is its anisotropic spectral energy transfer with respect to the background magnetic field.  As a result, small-scale plasma fluctuations populate predominantly field-perpendicular wavevectors and turbulence is dominated by the shear-Alfv\'en modes \cite{goldreich_s95}. In the linear case these modes have the dispersion relation $\omega=k_\| v_A$, where $k_\|$ is the field-parallel wavenumber with respect to the background magnetic field ${\bf B}_0$, $v_A=B_0/\sqrt{4\pi \rho}$ is the Alfv\'en velocity, and $\rho$ is the fluid density.   

At scales smaller than the so-called kinetic-Alfv\'en dispersion scale (ion acoustic radius if the electron temperature exceeds the ion temperature, or ion gyroradius otherwise) the assumptions of one-fluid MHD breaks down and the nature of turbulence changes. At such sub-proton scales, the  shear-Alfv\'en cascade transforms into the cascade of strongly anisotropic kinetic-Alfv\'en modes with a different linearized dispersion relation $\omega\propto k_\|k_\perp$. Kinetic Alfv\'en turbulence attracts considerable interest due to its importance for solar wind heating, magnetic reconnection in a variety of astrophysical systems, and laboratory experiments with strongly magnetized plasmas \cite[e.g.,][]{biskamp_etal1999,cho_l2004,kiyani_etal2009,chandran_etal10,kletzing_etal2010,howes2010,salem_etal2012}. 
Such turbulence has been understood to a much lesser extent compared to MHD turbulence. 

In this contribution we address the spectrum and structure of kinetic Alfv\'en turbulence. Our consideration is, in part,  motivated by measurements of small-scale fluctuations in the solar wind. Although significant scatter exists among the reported data \cite{smith_etal2006}, recent observations suggest that magnetic fluctuations at sub-proton scales have the Fourier energy spectrum close to or possibly steeper than  $k^{-2.8}$ \cite[e.g.,][]{alexandrova_etal2009,sahraoui_etal2009,chen_etal2010,alexandrova_etal2011,salem_etal2012}. The nature of such fluctuations is unclear as they are not described by existing models of either kinetic-Alfv\'en  or electron magnetohydrodynamic turbulence, which predict the scaling $k^{-7/3}$. Recently proposed explanations include significant steepening of the spectrum due to Landau damping, presence of weak turbulence, wave-particle scattering, etc.  \cite[e.g.,][]{rudakov_etal2011,howes_etal2011b}. 

In this paper we analyze small-scale kinetic-Alfv\'en turbulence using a two-fluid plasma description, which by its nature, does not take into account Landau damping and other wave-particle interactions. We found that the steeper than $k^{-7/3}$ energy spectrum persists in this case, in the form closely resembling the solar wind observations and the results of existing kinetic simulations. Our results indicate that the power-law energy spectrum of strong  kinetic-Alfv\'en turbulence in not an artifact of significant dissipation or non-universality, but rather an inherent property of nonlinear plasma dynamics. To describe this spectrum we propose a new model that assumes that the magnetic and density fluctuations tend to spontaneously organize into two-dimensional structures thus leading to strong spatial and temporal intermittency of turbulence. This happens as a consequence of kinetic-Alfv\'en dynamics that combines nonlinear striation in the field-perpendicular direction and linear spreading in the field-parallel direction. Our model predicts that the energy spectrum of strong kinetic-Alfv\'en turbulence has the Fourier scaling $E(k_\perp)\propto k_\perp^{-8/3}$, which is in good  agreement with our numerical findings and may explain the solar wind data. \\

{\em Kinetic Alfv\'en model.}---Let us assume that a uniform background magnetic field (the guide field) is strong compared to magnetic fluctuations, and we are interested in frequencies smaller than the ion gyrofrequency.  Since the electrons are strongly magnetized and their thermal speed exceeds the Alfv\'en speed, an isothermal fluid description is possible for the electrons~\footnote{This condition applies for a collisionless plasma. When collisions cannot be neglected, the electron fluid is isothermal if the electron diffusion time in the field-parallel direction $\tau_{diff}\sim 1/(k_\|^2 v_{Te}^2\tau_{coll})$ is less than the inverse frequencies of interest.}. The electrons are advected across the guide field by the ``E cross B'' drift, ${\bf v}_\perp= c{\bf E}\times {\bf B_0}/B_0^2$ while their field-parallel motion is related to the current $J_\|= - e n_e v_{e \|}$, and the ion parallel motion can be neglected. For simplicity, we start with the case when the thermal plasma energy is small compared to the magnetic energy (small plasma ``beta"). In this case the field-parallel fluctuations of the magnetic field can be neglected, while its  field-perpendicular component is expressed through the flux function ${\bf b}_\perp = {\hat z} \times \nabla \psi$, so that $J_\|=(c/4\pi)\nabla_\perp \times {\bf b}_\perp= (c/4\pi)\nabla_\perp^2 \psi$.  The flux function is the field-parallel component of the vector potential, $\psi= - A_z$. 

The field-parallel force balance in the electron momentum equation gives $- T_e \nabla_\| n_e -n_0 e{\bf E}_\|= 0$, where the electric field is ${\bf E}=-\nabla \phi -(1/c)\partial_t {\bf A}$. Supplementing this equation with the electron continuity equation, one obtains the system for the fluctuating parts of magnetic and density fields:
\begin{eqnarray}
& \frac{1}{c}\frac{\partial}{\partial t} {\psi}-\nabla_\| {\phi} +\frac{T_e}{n_0 e}\nabla_\|{n_e}= 0, 
\label{psi}\\
& \frac{\partial}{\partial t} { n_e}  - \frac{c}{B_0}\nabla {\phi}\times {\hat z}\cdot \nabla {n_e}- \frac{1}{e}\nabla_\| {J}_\|=0, 
\label{n}
\end{eqnarray}
These equations have been derived and studied in many works, e.g., \cite{hazeltine1983,scott_hd1985,camargo_etal1996,terry_mf2001,schekochihin_etal2009,smith_t2011}. The field-parallel gradient in these equations is the gradient along the {\em local} magnetic field, that is,
\begin{eqnarray}
\nabla_\|=\nabla_z + \frac{1}{B_0}{\hat z} \times \nabla {\psi} \cdot \nabla \,.
\label{nabla}  
\end{eqnarray}
We will assume that the fluctuations are anisotropic with respect to the local magnetic field, so that the so-called critical balance between the linear and nonlinear terms is satisfied, $\nabla_z \sim  (1/B_0){\hat z} \times \nabla { \psi} \cdot \nabla$, \cite[e.g.,][]{goldreich_s95,cho_l2004,howes_etal2011b,tenbarge_h2012}. This is the case of strong turbulence that we consider in this paper.  Equations (\ref{psi}, \ref{n}) are therefore essentially nonlinear and three-dimensional. 

We still need to specify the electric potential $\phi$ in the system (\ref{psi}, \ref{n}). We are interested in the dispersive kinetic-Alfv\'en waves, that is, we consider the scales smaller than the ion-acoustic scale $k_\perp \rho_s \gg 1$ or the ion gyroscale if $T_e\sim T_i$; here  $\rho_s=C_{s}/\Omega_{i}$, $C_s=(T_e/m_i)^{1/2}$ is the ion acoustic speed, and $\Omega_{i}$ is the ion gyrofrequency.  Below the ion gyroscale the ions are not magnetized, and since we are interested in frequencies smaller than $kv_{Ti}$ ($v_{Ti}=(T_i/m_i)^{1/2}$ is the ion thermal speed), we have for the ion density fluctuations $n_i=- e \phi n_0/T_i$.   
In this case, the quasi-neutrality condition $n_i = n_e$ ensures that the second (advection) term in Eq.~(\ref{n}) vanishes, while in Eq.~(\ref{psi}) the electric potential modifies the density term: $\nabla_\| \phi =-(T_i /n_0 e) \nabla_\| n_e$~\footnote{If $T_e \gg T_i$, one needs to consider separately the scales above and below the ion gyroscale $\rho_i=v_{Ti}/\Omega_i$. Above the ion gyroscale (but below $\rho_s$), one can demonstrate that $\phi_k\sim (v_A/c)\psi_k/(k_\perp \rho_s)$, e.g., \cite{terry_mf2001}; the electric potential is small and the $\phi$-containing terms can be neglected in Eqs.~(\ref{psi}, \ref{n}).}. 

Let us introduce the normalized electron density ${\tilde n}=(1+T_i/T_e)^{1/2}(C_s/v_A) n_e/n_0$, magnetic flux function ${\tilde \psi}=(C_s/c)e A_z/T_e$, and the electric potential ${\tilde \phi}=(C_s/v_A)e\phi/T_e$. 
We normalize the spatial scales to the ion-acoustic scale $\rho_s$, and the time scale to $(\rho_s/v_A)(1+T_i/T_e)^{-1/2}$. In what follows we will use only the normalized variables and omit the over-tilde sign. Then we obtain that the magnetic and density fields in the kinetic Alfv\'en regime are described by the system:
\begin{eqnarray}
& \partial_t {\psi} + \nabla_\|{n}= 0, 
\label{psimod}\\
& \partial_t {n} -\nabla_\| \nabla_\perp^2 \psi=0,  
\label{nmod} 
\end{eqnarray}
where $\nabla_\|=\nabla_z+ {\hat z} \times \nabla {\psi} \cdot \nabla_\perp$. The presented ideal system conserves the total energy $E$ and the cross-correlation $H$,
\begin{eqnarray}
& E=\int \left( |\nabla \psi|^2 + n^2 \right)d^3 x, 
\label{energy}\\
& H=\int \psi n d^3 x .
\end{eqnarray}
The system (\ref{psimod},\ref{nmod}) possesses linear waves, $n_k\propto \psi_k \propto \exp(-i\omega t +i{\bf k}{\bf x})$. The linearization is done by neglecting the second term in the right-hand side of Eq.~(\ref{nabla}), which gives the dispersion relation for the kinetic Alfv\'en waves: 
\begin{eqnarray}
\omega= k_\| k_\perp .
\label{dispersion}
\end{eqnarray}
The linear modes are characterized by the equipartition of density and magnetic fluctuations, $n_k=\pm k_\perp \psi_k$. 

To conclude this section we make two important comments. First, a similar consideration can be conducted  without the assumption of small plasma beta. In this case the field-parallel fluctuations of the magnetic field should be taken into account in the derivation of (\ref{psi}, \ref{n})   \cite[e.g.,][]{howes_etal2006,schekochihin_etal2009}. The resulting system however has the structure identical to our system (\ref{psimod}, \ref{nmod}) and it can be reduced to system (\ref{psimod}, \ref{nmod}) by appropriate normalization of the variables. The value of beta is therefore not essential for our discussion of scaling properties of kinetic Alfv\'en turbulence. Second, we are interested in the processes slower than the ion gyrofrequency, implying $k_\|\ll 1$. However, equations (\ref{psimod}, \ref{nmod}) admit a rescaling $\partial/\partial t \to \epsilon \partial/\partial t$, $\nabla_z \to \epsilon \nabla_z$, ${n}\to \epsilon {n}$, and ${\psi} \to \epsilon { \psi}$ with arbitrary $\epsilon$, which preserves the critical balance.  This reflects the fact that being derived in the limit of infinitely large electron gyrofrequency  Eqs.~(\ref{psimod}, \ref{nmod}) lack any frequency scale. We may therefore always rescale the fields in these equations to satisfy $k_\|\sim n \sim \psi \sim 1$. Such rescaling will be used in our numerical simulations below.\\

{\em Kinetic Alfv\'en turbulence.}---The scaling of strong kinetic Alfv\'en turbulence was addressed in a number of works, \cite[e.g.,][]{howes_etal2008b,schekochihin_etal2009}, see also \cite{biskamp_etal1999,ng03,cho_l2004,cho_l2009}. It was argued that in strong turbulence, the critical balance condition which ensures that both linear and nonlinear terms in (\ref{nabla}) are of the same order, should be satisfied at all scales. Denote $n_\lambda$ and $\psi_\lambda$ the typical (rms) fluctuations at the field-perpendicular scale $\lambda$, and $l$ the corresponding field-parallel scale of those fluctuations. Balancing linear and nonlinear terms in (\ref{nabla}) then gives $l\sim \lambda^2/\psi_\lambda$, in which case the time of nonlinear interaction is comparable to the linear time (\ref{dispersion}), $\tau \sim 1/\omega \sim  l\lambda \sim \lambda^3/ \psi_\lambda$. Besides, we estimate from (\ref{psimod}, \ref{nmod}) that $n_\lambda \sim \psi_\lambda /\lambda$. The energy associated with the scale $\lambda$ can therefore be estimated as $E_\lambda\sim n_\lambda^2$, and the condition of constant energy flux in the turbulent cascade leads to $n_\lambda^2/\tau ={\rm const}$, which translates into the scaling for the turbulent fields $n_\lambda \sim \psi_\lambda/\lambda \sim \lambda^{2/3}$. The Fourier energy spectrum of strong kinetic Alfv\'en turbulence is then:
\begin{eqnarray} 
E_{KA}(k_\perp)\,dk_\perp \propto k_\perp^{-7/3}\,dk_\perp .
\label{spectrum}
\end{eqnarray}
As we discussed in the introduction, there is a puzzling disagreement of this scaling with the solar wind observations, where a spectrum closer to $-2.8$ is observed.  

To address this issue we have conducted numerical simulations of system (\ref{psimod},\ref{nmod}). Our results produce a turbulent spectrum that is {\em different} from (\ref{spectrum}), and quite close to the observational data. Since our system does not include Landau damping and it is driven in the regime of strong turbulence, we propose that the observed scaling is not an artifact of non-universal or dissipative effects, rather, it is an inherent property of nonlinear turbulent dynamics. We then propose a model of kinetic Alfv\'en turbulence, which predicts that the energy spectrum should scale as $E(k_\perp)\propto k_\perp^{-8/3}$, in good  agreement with our numerical results.   \\

\begin{figure}[tbp!]
\includegraphics[width=1.\columnwidth]{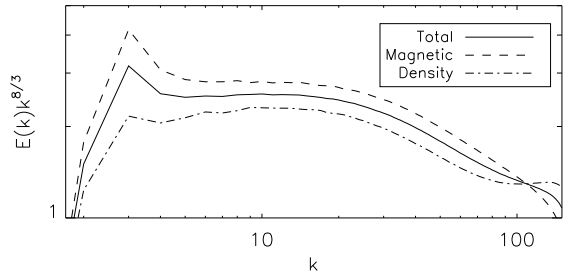}
\caption{\label{fig:spectrum} The energy spectrum of strong kinetic Alfv\'en turbulence below ion acoustic scale (ion gyroscale).}
\end{figure}

{\em Numerical simulations of kinetic Alfv\'en turbulence}---We supplement the system  (\ref{psimod},\ref{nmod}) by a driving force and by small dissipation terms as follows:
\begin{eqnarray}
& \partial_t {\psi} + \nabla_\|{n}= \eta \nabla_\perp^2 \psi + f, 
\label{eq1}\\
& \partial_t {n} - \nabla_\| \nabla_\perp^2 \psi = \nu \nabla_\perp^2 n. 
\label{eq2}
\end{eqnarray}
The force mimics energy supply from large-scale motion, while the dissipation terms (normalized plasma resistivity $\eta$ and electron diffusivity $\nu$) remove the energy at small scales; the dissipation terms are also needed to ensure numerical stability of the code.  We solve these equations on a triply periodic cubic domain ($L^3$, $L=1$) using standard pseudo-spectral methods. The random force  $f$ is applied in Fourier 
space at wavenumbers $2\pi/L \leq k_{\perp} \leq 2 (2\pi/L)$, $k_\|= 2\pi/L$.  
The Fourier coefficients outside the above range are zero and inside that range are 
Gaussian random numbers with amplitudes chosen so that $|\nabla \psi|_{rms}\sim 1$. The individual random values are refreshed independently on average every $\tau=0.1  L/(2\pi |\nabla \psi|_{rms})$. We choose $\nu=\eta=0.01$. The strength of the nonlinear term relative to the dissipation term is then measured by the parameter $R=\psi_{rms}/(L \nu) \sim n_{rms}/\nu$, which plays a role of the Reynolds number in this system. 

We use numerical resolution of $512^3$ collocation points. The initial conditions are imported from a  steady state snapshot obtained on $256^3$ points. The system is then evolved until a new steady state is reached. The $512^3$ simulations are run for about 35 large-scale dynamical times. The presented results correspond to statistical averages over approximately 60 last snapshots corresponding to about 15 dynamical times. Note that compared to the MHD equations where $\omega \sim k_\|$, the kinetic Alfv\'en equations require significantly shorter time steps to accommodate high  frequencies (\ref{dispersion}), leading to tremendous increase in computational effort. In this respect the fluid model (\ref{eq1},\ref{eq2}) allows one to access the inertial intervals and averaging times currently unachievable in kinetic or gyrokinetic simulations. Fig.(\ref{fig:spectrum}) shows the energy spectrum of kinetic Alfv\'en turbulence. The spectrum is steeper than $-7/3$ and close to $-8/3$. \\

\begin{figure}
\includegraphics[width=\columnwidth]{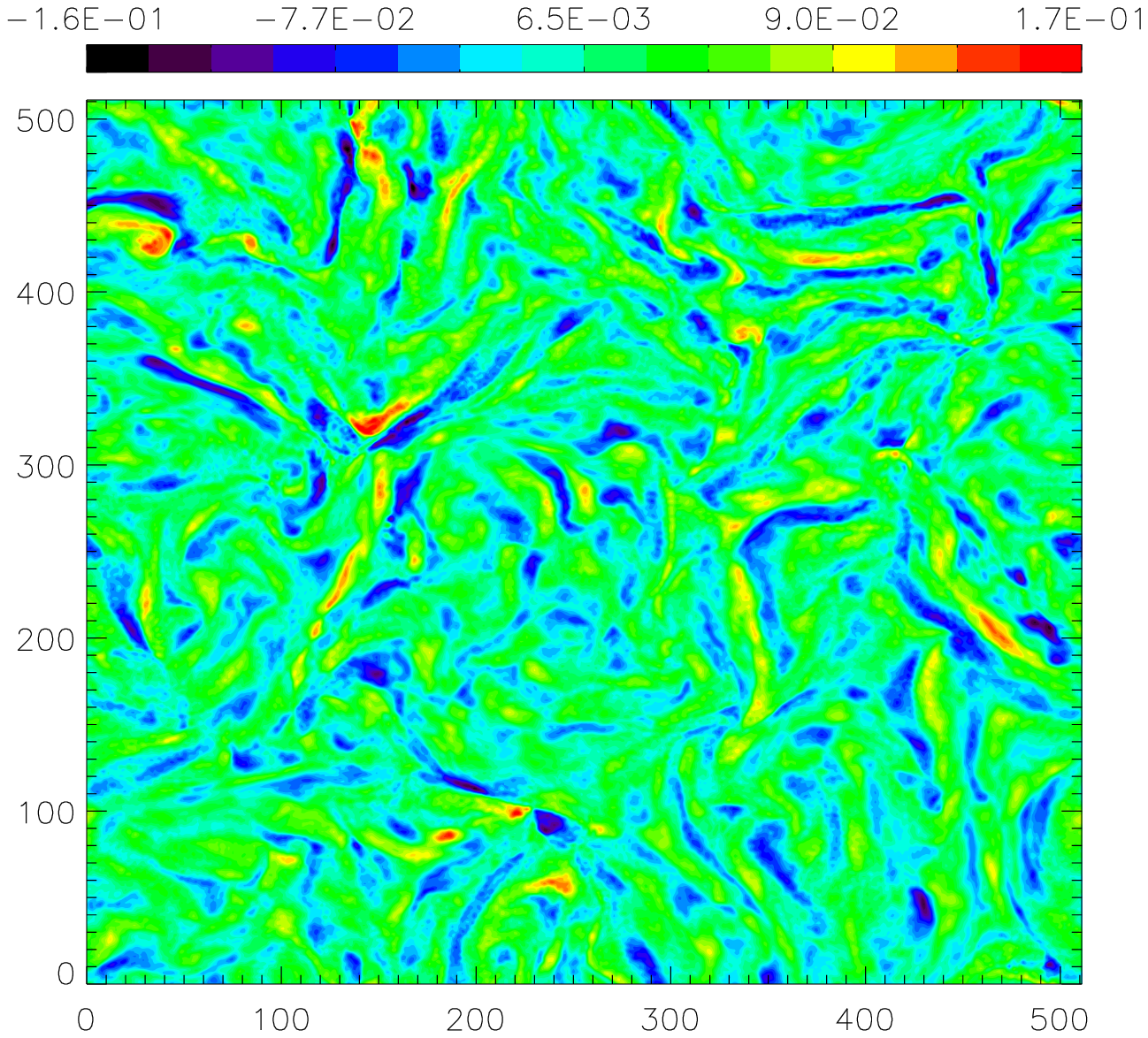}
\includegraphics[width=\columnwidth]{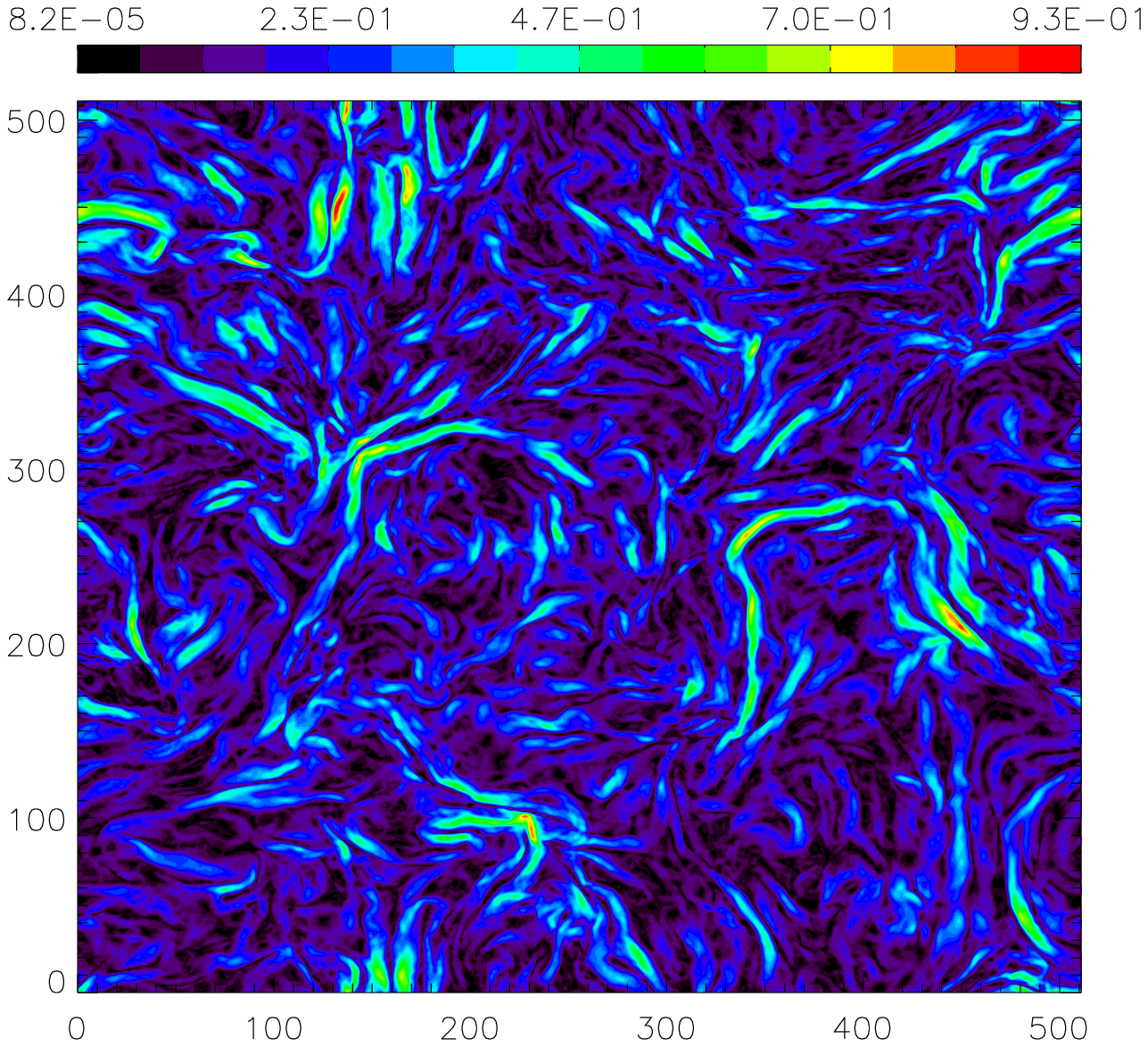}
\caption{\label{fig:structures} Density fluctuations (top) and amplitude of magnetic-field fluctuations (bottom) in a field-perpendicular cross section of a simulation domain. The large-scale harmonics with $k<2\pi/12$ have been filtered out; the plot thus represents fluctuations in the inertial interval, see Fig.~(\ref{fig:spectrum}). The plot suggests that both density and magnetic fluctuations are concentrated along two-dimensional structures. Corresponding field-parallel cross sections (not shown here) are consistent with this picture.}
\end{figure}

{\em A model for kinetic Alfv\'en turbulence}---To understand the observed energy spectrum, let us discuss some characteristic properties of the dynamics described by the system (\ref{psimod}, \ref{nmod}). First, consider the effect of the nonlinear terms (for that we can assume $k_\|=0$). The nonlinear term in Eq.~(\ref{eq1}) can be rewritten as $\nabla n \times {\hat z}\cdot \nabla \psi$, implying that $\psi$ is advected in the field-perpendicular direction with velocity $\nabla n \times {\hat z}$. The field $\psi$ thus gets striated, developing gradients aligned with the gradients of $n$. This suggests that the magnetic field ($\nabla \psi$) tends to concentrate in 2D structures. Let us now see what happens to those structures if the linear terms come into play (for that we can assume a non-zero $k_\|$). The linear terms in (\ref{psimod},\ref{nmod}) tend to smear or break the initial perturbation into wave packets propagating in opposite directions along the local magnetic field, such that $n \sim |\nabla \psi|$ inside those packets.  Thus density tends to get in equipartition with the magnetic field and to concentrate in 2D structures as well. We thus expect that as a result of nonlinear striation and linear propagation, both the density and the magnetic fluctuations get organized in highly intermittent, two-dimensional structures or sheets, elongated in $\hat z$ direction. This is indeed consistent with our  numerical observations presented in Fig~(\ref{fig:structures}).     

We therefore assume that  essential nonlinear interaction and energy cascade take place at such 2D structures. Following a standard procedure \cite[e.g.,][]{frisch1995}, consider turbulent fluctuations of field-perpendicular size $\lambda$. Since such fluctuations cover 2D sheets, they occupy the volume fraction $p_\lambda \propto \lambda $. The energy density of such fluctuations therefore scales as $E_\lambda \propto n_\lambda^2 p_\lambda$. The energy cascade time is estimated as before [cf. discussion preceding (\ref{spectrum})], $\tau \sim 1/\omega \sim l\lambda \sim \lambda^3/\psi_\lambda \sim \lambda^2/n_\lambda $, and the condition of constant energy flux reads $E_\lambda /\tau = {\rm const}$, which gives $n_\lambda \propto  \lambda^{1/3}$. The scaling of the energy is then $E_\lambda \propto \lambda^{5/3}$, and the Fourier energy spectrum scales as
\begin{eqnarray}
E(k_\perp ) \,dk_\perp \propto k_\perp^{-8/3}\,dk_\perp .
\label{kaw_spectrum}
\end{eqnarray} 
This spectrum of kinetic Alfv\'en turbulence is in excellent agreement with the numerical observation in Fig.~(\ref{fig:spectrum}), and it is the main result of our work. It provides a plausible explanation for the solar wind measurements, e.g.~\cite{chen_etal2010,alexandrova_etal2011}. Balancing the linear wave frequency $\omega \sim 1/(l \lambda )$ with the inverse nonlinear interaction time $1/\tau \sim n_\lambda/\lambda^2 \sim \lambda^{-5/3}$ we further derive the anisotropy of the turbulent fluctuations with respect to the (local) large-scale magnetic field: $l\sim \lambda^{2/3}$. If we formally introduce the local field-parallel wave number as $k_\|\sim 1/l$, then the ``field-parallel energy spectrum'' corresponding to (\ref{kaw_spectrum}) is $E(k_\|)\,dk_\| \propto k_\|^{-7/2}\,dk_\| $. This is also consistent with the solar wind measurements \cite{chen_etal2010}.  As for the spectrum of the electric field, it is a factor of $k_\perp^2$ flatter than (\ref{kaw_spectrum}), $E_E\propto k_\perp^{-2/3}$ \footnote{In the case $T_e \gg T_i$, in the interval of scales between $\rho_s$ and $\rho_i$, the electric spectrum is a factor of $k_\perp^2$ steeper than (\ref{kaw_spectrum})\cite{terry_mf2001}, $E_E\propto k_\perp^{-14/3}$.}.\\

{\em Discussion}---We have proposed a model for kinetic Alfv\'en turbulence below the dispersion scale (ion-acoustic scale). Based on numerical simulations of the fluid equations (\ref{eq1},\ref{eq2}) and on analytic modeling we propose that the energy spectrum of such turbulence scales as $k_\perp^{-8/3}$, meaning that both magnetic and density fluctuations should have the same Fourier spectrum. This result is also consistent with in situ observations of the sub-proton solar wind fluctuations, and with the results of the gyro-kinetic simulations where the spectra close to $k_\perp^{-2.8}$ are observed \cite{chen_etal2010,alexandrova_etal2011,howes_etal2011b}.

Our model is complementary to the previously proposed explanations invoking Landau damping, the presence of weak kinetic Alfv\'en turbulence, effects of wave-particle scattering, etc. \cite[e.g.,][]{rudakov_etal2011,howes_etal2011b}. These explanations are interesting and the effects they point out may indeed affect a turbulent cascade. The  difference of our approach (as compared to the gyrokinetic numerical studies, for example) is that it does not take into account Landau damping and it allows us to drive turbulence in a strongly coupled state. With those ``spoilers'' removed, the observed turbulent spectrum $k_\perp^{-8/3}$ is expected to arise from  the nonlinear interaction, similar to the Kolmogorov spectrum of hydrodynamic turbulence. This also implies that dissipative effects may be less important at the subproton scales than was previously thought.  
  
Our explanation points out to an interesting property of the kinetic-Alfv\'en nonlinear dynamics, which  tend to concentrate magnetic and density fluctuations at two-dimensional structures. This leads to strong spatio-temporal intermittency in the field distributions. In this work we have studied only the second-order statistics of the fluctuating fields (expressed through energy spectra); we plan to present more detailed discussion of the sub-proton turbulence elsewhere.

Finally, we note that a system formally similar to our system of equations (\ref{eq1}), (\ref{eq2}) also appears in the limit of strong guide magnetic field in the so-called electron MHD, where the ions are assumed to be immobile. The wave modes in this case correspond to the so-called whistler waves and belong to a different branch of plasma dispersion relations (this branch continues into a magnetohydrodynamic mode or compressional Alfv\'en mode in the limit of low $k$, while the kinetic Alfv\'en mode continues into the shear Alfv\'en mode, \cite[e.g.,][]{sovinec2009}).  One therefore expects that our simulations of turbulence based on the system (\ref{eq1},\ref{eq2}) should  be relevant for  electron MHD in the limit of strong guide field. Previous simulations of electron MHD were conducted mostly for two-dimensional cases and/or for relatively weak guide fields and/or decaying cases, and the spectrum close to $k_\perp^{-7/3}$ was observed \cite[e.g.,][]{biskamp_etal1999,cho_l2004,ng03,dastgeer_z2003,cho_l2009}. Our results suggest that in the limit of strong guide field, the spectrum of driven strong electron-MHD turbulence may be modified as well.

\begin{acknowledgments}
We are grateful to Christopher Chen for useful discussions. This work was supported   
by the US DoE grants DE-FG02-07ER54932, DE-SC0003888, DE-SC0001794, the NSF Grant PHY-0903872, the NSF/DOE Grant AGS-1003451, and the NSF Center for Magnetic Self-organization in Laboratory and Astrophysical Plasmas at U. Wisconsin-Madison. High Performance Computing resources were
provided by the Texas Advanced Computing Center (TACC) at the
University of Texas at Austin under the NSF-Teragrid Project
TG-PHY080013N. 
\end{acknowledgments}


\end{document}